\documentclass[epj]{svjour}

\def\be{\begin{equation}} 
\def\ee{\end{equation}} 

\def\bea{\begin{eqnarray}}
\def\ena{\end{eqnarray}}

\RequirePackage{graphicx}

\usepackage{cite} 
\usepackage{amsmath}
\usepackage{hyperref}
\usepackage{amsmath,amssymb}
\usepackage{xcolor}

\usepackage{graphics}

\begin{document}
\title{Relativistic strange quark stars in Lovelock gravity}

\author{
Grigoris Panotopoulos \inst{1} 
\thanks{E-mail: \href{mailto:grigorios.panotopoulos@tecnico.ulisboa.pt}{\nolinkurl{grigorios.panotopoulos@tecnico.ulisboa.pt}} }
\and
\'Angel Rinc\'on \inst{2}
\thanks{E-mail: \href{mailto:angel.rincon@pucv.cl}{\nolinkurl{angel.rincon@pucv.cl}} }
}                     
%
%
\institute{ 
Centro de Astrof{\'i}sica e Gravita{\c c}{\~a}o, Departamento de F{\'i}sica, Instituto Superior T\'ecnico-IST,
\\
Universidade de Lisboa-UL, Av. Rovisco Pais, 1049-001 Lisboa, Portugal
\and
Instituto de F{\'i}sica, Pontificia Universidad Cat\'olica de Valpara{\'i}so, Avenida Brasil 2950, Casilla 4059, Valpara{\'i}so, Chile
}

\date{Received: date / Revised version: date}
%
\abstract{
We study relativistic non-rotating stars in the framework of Lovelock gravity. In particular, we consider the Gauss-Bonnet term in a five-dimensional spacetime, and we investigate the impact of the Gauss-Bonnet parameter on properties of the stars, both isotropic and anisotropic. For matter inside the star, we assume a relativistic gas of de-confined massless quarks. We integrate the modified Tolman-Oppenheimer-Volkoff equations numerically, and we obtain the mass-to-radius profile, the compactness of the star as well as the gravitational red-shift for several values of the Gauss-Bonnet parameter. The maximum star mass and radius are also reported.
}

\maketitle

\section{Introduction}

Although our observable Universe is clearly four-dimensional, the question "How many dimensions are there?" is one of the fundamental questions High Energy Physics tries to answer. Kaluza-Klein theories \cite{kaluza,klein}, supergravity \cite{nilles} and Superstring/M-Theory \cite{ST1,ST2} have pushed forward the idea that extra spatial dimensions may exist. In more than four dimensions higher order curvature terms are natural in Lovelock theory \cite{Lovelock}, and also higher order curvature corrections appear in the low-energy effective equations of Superstring Theory \cite{ramgoolam}. 

The amount of published works in the literature reveals that Lovelock gravity is an exciting and active field. Black hole physics \cite{BH1,BH2,BH3,BH4,BH5,BH6,BH7}, cosmological solutions \cite{cosmo1,cosmo2,cosmo3,cosmo4} and holographic superconductors exploiting the gauge/gravity duality \cite{holography} are some of the areas that have been explored in the framework of Lovelock gravity. However, the astrophysical implications of Lovelock theory should also be investigated. Compact objects \cite{textbook,review1,review2}, such as white dwarfs and neutron stars, are the final fate of stars, and comprise excellent cosmic laboratories to study, test and constrain new physics and/or alternative theories of gravity under extreme conditions that cannot be reached in Earth-based experiments. It is well-known that the properties of compact objects, such as mass and radius, depend both on the equation-of-state (EoS) of ultra-dense matter and on the underlying theory of gravity. 

A couple of years after the discovery of the neutron by James Chadwick \cite{neutron1,neutron2},
neutron stars were predicted to exist by Baade and Zwicky \cite{baade}. Indeed, several decades after that, the discoveries of pulsars in the Crab and Vela supernova remnants \cite{chamel} led to their identification as neutron stars just one year after the discovery of pulsars in 1967 \cite{pulsars}. On the other hand quark matter is by assumption absolutely stable, and as such it may be the true ground state of hadronic matter \cite{witten,farhi}. Therefore a new class of hypothetical compact objects has been postulated to exist. These compact objects could serve as an alternative to neutron stars, and they might offer us a plausible explanation of the puzzling observation of some super-luminous supernovae \cite{SL1,SL2}, which occur in about one out of every 1000 supernovae explosions, and which are more than 100 times more luminous than regular supernovae. The compact objects called "strange quark stars" \cite{SS1,SS2,SS3,SS4,SS5,SS6}, since they are a much more stable configuration in comparison with neutron stars, may explain the origin of the huge amount of energy released in super-luminous supernovae. 

Nowadays there is indeed some evidence that strange quark stars might exist. In particular, the recent discovery of very compact objects, that is celestial bodies with very high densities, such as the millisecond pulsars SAX J 1808.4-3658 and RXJ185635-3754, the X-ray burster 4U 1820-30, the X-ray pulsar Her X-1, and the X-ray source PSR 0943+10, are currently among the best candidates \cite{lattimer}. What is more, the recently launched NASA mission NICER, designed primarily to observe thermal X-rays emitted by several millisecond pulsars, may help answering this question \cite{nicer}.

Previously relativistic spherically symmetric stars were studied in Lovelock gravity, and some exact solutions of the modified Tolman-Oppenheimer-Volkoff (TOV) equations were obtained \cite{paper1,paper2,paper3}. In the present work, to the best of our knowledge, for the first time we investigate properties of relativistic stars considering a linear equation-of-state describing quark matter. Our work is organized as follows: In the next section we present the equation-of-state for a relativistic gas of de-confined quarks inside the star as well as the modified TOV equations. In section 3 we integrate the structure equations numerically to obtain the mass-to-radius profile, the compactness and the gravitational red-shift, and we discuss our numerical results. In the fourth section we study the impact of the Gauss-Bonnet parameter on the internal solution of anisotropic objects. We finish concluding our work in the last section. We adopt the mostly negative metric signature $(+,-,-,-,-)$, and we work in units where $\hbar=c=8 \pi G_5 = 1$.

\section{Hydrodynamical equilibrium in Lovelock gravity}

\subsection{Field equations of Einstein-Gauss-Bonnet gravity}

In this section we shall be using small letters for the four-dimensional Einstein's theory and capital letters for the five-dimensional gravity. 

In the framework of Lovelock gravity \cite{Lovelock} we consider the model described by the action
\be
S = \frac{1}{2} \int \mathrm{d}^5x \sqrt{-g_5} (R_5 + \alpha L_{GB}) + S_M
\ee
where $S_M$ is the matter contribution corresponding to a perfect fluid with stress-energy tensor $T_{MN}$, $R_5$ is the five-dimensional Ricci scalar, $g_5$ is the determinant of the five-dimensional metric tensor $g_{A B}$, $\alpha$ is the Gauss-Bonnet parameter, while the Gauss-Bonnet Lagrangian is given by \cite{paper1,paper2}
\be
L_{GB} = R_5^2 - 4 R_{AB} R^{AB} + R_{ABCD} R^{ABCD}
\ee
with $R_{AB}$ and $R_{ABCD}$ are the five-dimensional Ricci and Riemann tensor, respectively. 
Varying with respect to the metric tensor we obtain the field equations
\be
G_{MN} + \alpha H_{MN} = -T_{MN}
\ee
where $G_{MN}$ is the five-dimensional Einstein tensor, while the new tensor $H_{MN}$, absent in Einstein's theory, is given by \cite{paper1,paper2}
\be
H_{MN} = - \frac{1}{2} L_{GB} g_{MN} + 2 \Bigl( R R_{MN} + R_M^{ABC} R_{NABC} \\
- 2 R_{MC} R_N^C -2 R^{CD} R_{MCND} \Bigl)
\ee

\subsection{Equation-of-state}

For strange matter we shall consider a linear EoS corresponding to the simplest version of the MIT bag model \cite{MIT1,MIT2}, which describes 
a relativistic gas of massless de-confined quarks
\be
p = \frac{1}{4} (\rho - 5B)
\ee
with $B$ being the bag constant. Notice that contrary to the four-dimensional version of the EoS where the numerical prefactor is $1/3$, here it is $1/4$ due to the extra dimension. Also the surface energy density here becomes $\rho_s=5B$ instead of $4B$ in four dimensions.

The above EoS may be easily derived in the five-dimensional case as follows: On the one hand, at zero temperature, $T=0$, and if u, d and s quarks are taken to be massless, both the energy density $\rho$ and the pressure $p$ are given entirely in terms of the quark chemical potential $\mu$, and given the dimensions there is only one option in $1+4$ dimensions
\begin{equation}
p = a \mu^5 - B
\end{equation}
where we have to substract the bag constant, and $a$ is some numerical prefactor which does not need to be specified. On the other hand, from standard thermodynamics it is known that at $T=0$ the energy density is computed by
\begin{equation}
\rho = n \mu - p
\end{equation}
where the quark number density is given by $n=\partial p/\partial \mu$.
Therefore, the energy density is found to be
\begin{equation}
\rho = 4 a \mu^5 + B
\end{equation}
Finally, eliminating the chemical potential we may derive the EoS $p(\rho)$ given above.

Although refinements of the bag model exist in the literature \cite{refine1,refine2,refine3} (for the present state-of-the-art see the recent paper \cite{art}), the above analytical expression "radiation plus constant" has been employed in recent works, both in GR \cite{prototype,greg1,greg2,greg3}, and in R-squared gravity \cite{kokkotas1,kokkotas2,greg4}. Since the aim of this work is to investigate the impact of the Gauss-Bonnet parameter on properties of strange quark stars, we will consider here the MIT bag model despite its simplicity, as it suffices for our purposes.

\subsection{Structure equations}

First let us briefly review the standard TOV equations for 
relativistic stars in General Relativity (GR) \cite{GR}. The starting point 
is Einstein's field equations without a cosmological constant
\be
G_{\mu \nu} = R_{\mu \nu} - \frac{1}{2} R_4 g_{\mu \nu}  = -8 \pi G_N T_{\mu \nu}
\ee
where for matter we assume a perfect fluid with energy density $\rho$, pressure $p$ and some EoS $p(\rho)$, and $G_N$ is the usual Newton's constant in four dimensions. Assuming static spherically symmetric solutions of the form
\be
\mathrm{d}s^2 = e^{2 \nu(r)} \mathrm{d}t^2 -  \left(1 - \frac{2m(r)}{r} \right)^{-1}  \mathrm{d}r^2 - r^2 \mathrm{d} \Omega_2^2
\ee
where $\mathrm{d} \Omega_2^2$ is the usual line element of the unit two-dimensional sphere,
we obtain the Tolman-Oppenheimer-Volkoff equations using the $tt$, the $rr$ and the conservation of the energy-momentum tensor \cite{TOV1,TOV2}
\bea
m'(r) & = & 4 \pi r^2 \rho(r) \\
\nu'(r) & = & \frac{m(r) + 4 \pi p(r) r^3}{r^2 (1-2m(r)/r)} \\
p'(r) & = & - (p(r)+\rho(r)) \: \nu'(r)
\ena
setting the usual four-dimensional Newton's constant $G_N=1$. Upon matching with the exterior solution (where the energy-momentum tensor vanishes), which is no other than the Schwarzschild solution \cite{SBH}
\be
\mathrm{d}s^2 = \left(1-\frac{2M}{r}\right) \mathrm{d}t^2 - \left(1-\frac{2M}{r}\right)^{-1}  \mathrm{d}r^2 - r^2 \mathrm{d}\Omega_2^2
\ee
with $M$ being the mass of the star, the radius of the star $R$ is determined requiring that the pressure vanishes at the surface, $p(R) = 0$, while the mass of the star is then given by $M=m(R)$. Finally, the other metric function, $\nu(r)$, can be computed using the second equation together with the boundary condition 
\begin{align}
2 \nu(R)=\ln\left(1-\frac{2M}{R}\right).
\end{align}

Next, in the framework of the five-dimensional Lovelock gravity with a Gauss-Bonnet term we follow the same steps as in four-dimensional GR. We seek static spherically symmetric solutions of the form
\be
\mathrm{d}s^2 = e^{2 \nu(r)} \mathrm{d}t^2 - e^{2 \lambda(r)}  \mathrm{d}r^2 - r^2 \mathrm{d} \Omega_3^2
\ee
where $d \Omega_3^2$ is the line element of the unit three-dimensional sphere given by \cite{paper1,paper2}
\be
\mathrm{d} \Omega_3^2 = \mathrm{d} \theta^2 + \sin^2 \theta \mathrm{d} \phi^2 + \sin^2 \theta \sin^2 \phi \mathrm{d}\psi^2
\ee
First we need to know the exterior solution, $r > R$, where $T_{AB}=0$, which is the Lovelock version of the vacuum Schwarzschild solution, and for which the metric function is found to be \cite{deser} (the corresponding solutions with a non-vanishing cosmological constant have been obtained in \cite{cai1,cai2}, and they are reduced to the vacuum solution when the cosmological constant is set to zero)
\be
f(r) = 1 + \frac{r^2}{4 \alpha} \left [ 1 - \sqrt{1+\frac{16 M \alpha}{3 A_3 r^4}} \right ]
\ee
with $A_3=2 \pi^2$. Introducing the mass function 
\be
e^{-2 \lambda(r)} = 1 + \frac{r^2}{4 \alpha} \left [ 1 - \sqrt{1+\frac{16 m(r) \alpha}{3 A_3 r^4}} \right ]
\ee
we obtain for the interior solution ($r < R$) the modified TOV equations \cite{paper1,paper2}

\begin{align}
m'(r) & =  2 \pi^2 r^3 \rho(r) \\
p'(r) & =  - (p(r) + \rho(r)) \: \frac{\frac{p(r) r^3}{3} + r (1-e^{-2 \lambda})}{(r^2 + 4\alpha) e^{-2 \lambda} - 3 \alpha e^{-4 \lambda}} \\
\nu'(r) & =  - \frac{p'(r)}{p(r) + \rho(r)}
\end{align}

where the first equation is the five-dimensional version of the mass function
\be
m(r) = 2 \pi^2 \int_0^r \mathrm{d}x \: x^3 \rho(x)
\ee
while the third equation expresses the conservation of the stress-energy tensor, and remains the same as in four-dimensional GR. 

We integrate the structure equations numerically from the center $r=0$ to the surface of the star $r=R$ with the initial conditions 
\bea
m(0) & = & 0 \\
p(0) & = & p_c
\ena
where $p_c$ is the central pressure. Finally, to match the two solutions we impose the conditions on the surface of the star
\bea
p(R) & = & 0 \\
m(R) & = & M
\ena
which allow us to determine the radius and the mass of the star.

\begin{figure}[ht!]
\centering
\includegraphics[width=0.49\textwidth]{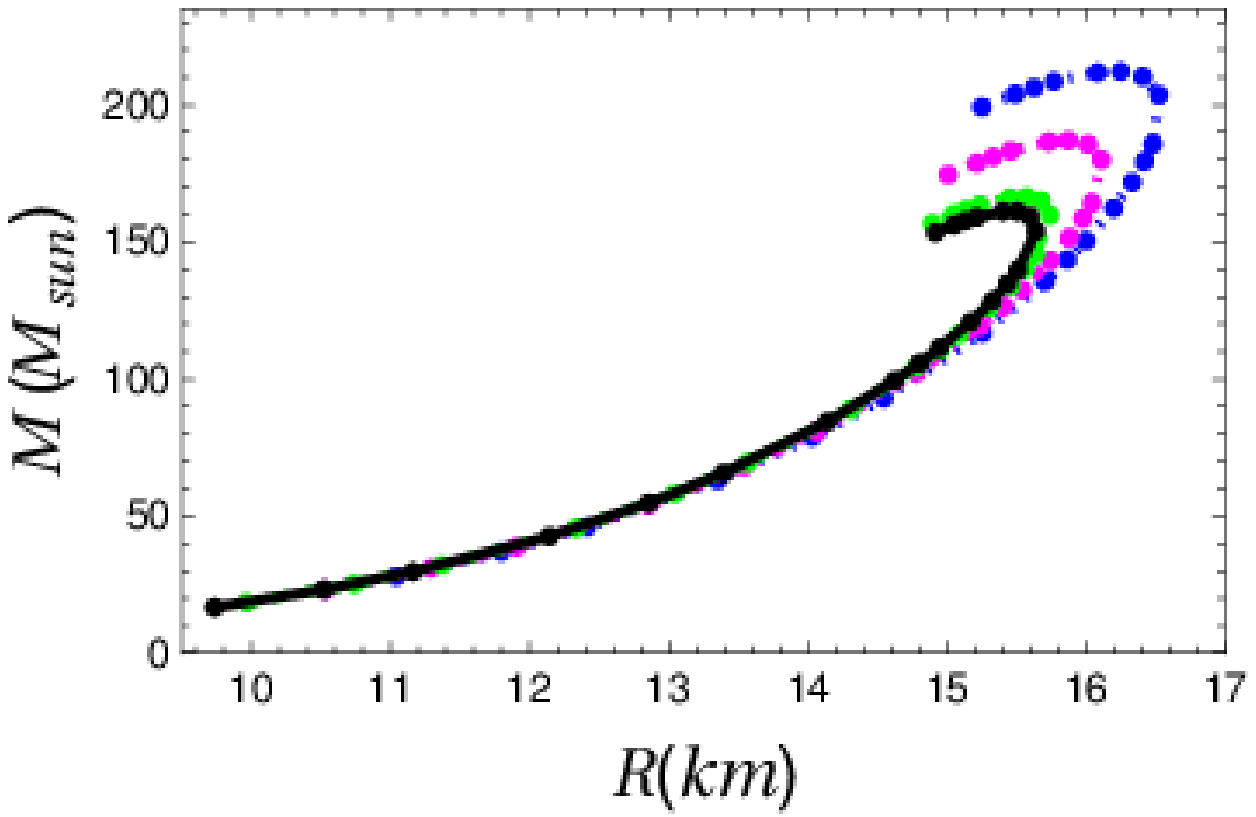} \ 
\includegraphics[width=0.49\textwidth]{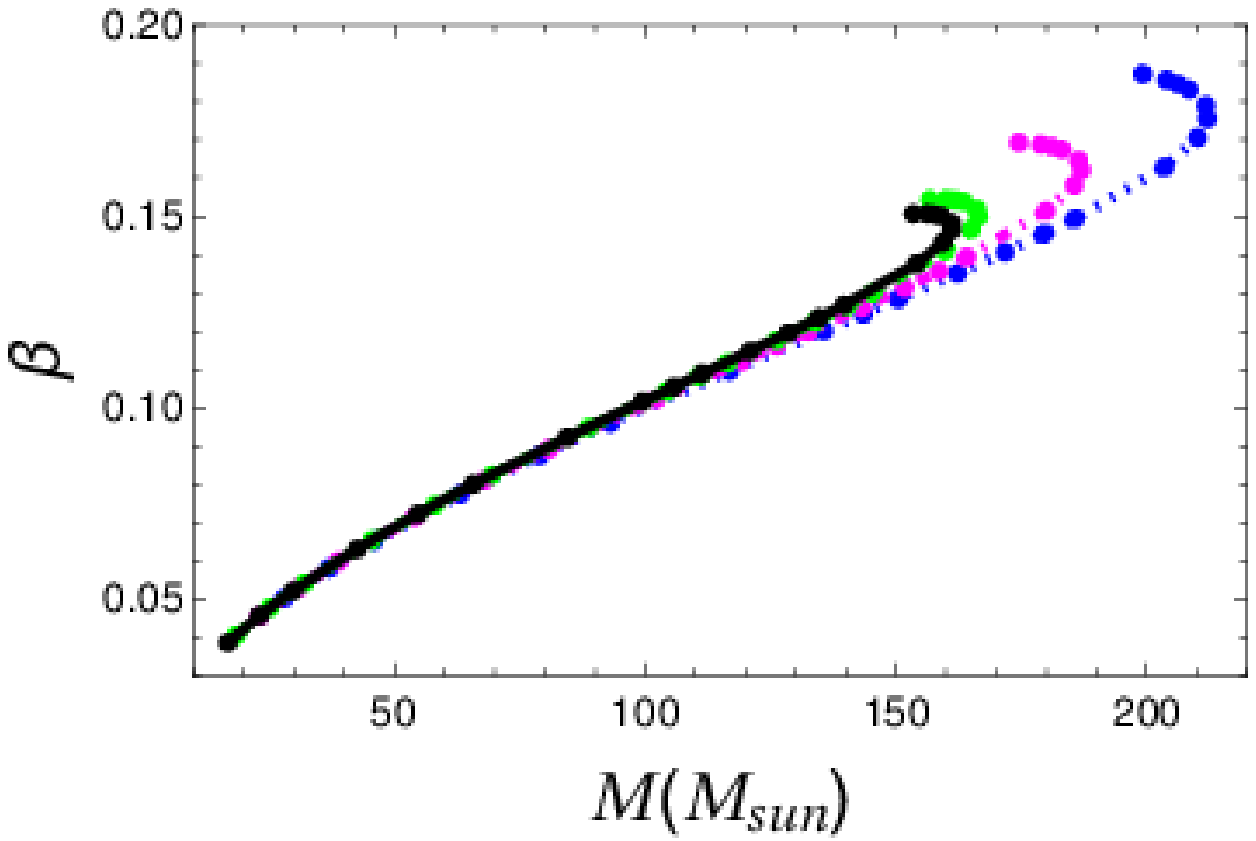}
\caption{{\bf Left panel:} 
Mass-to-radius profiles (mass in solar masses and radius in km) for three different values of the Guass-Bonnet parameter. The case $\alpha=0$ is also shown for comparison reasons. {\bf Right panel:} Compactness of the star $\beta=(6 \pi^2)^{-1} (M/R^2)$ vs mass of the star (in solar masses) for three different values of the Guass-Bonnet parameter. The case $\alpha=0$ is also shown for comparison reasons.
The color code is as follows: i) black curve for $\alpha=0$ (5D GR), ii) green curve for $\alpha=1$, iii) magenta curve for $\alpha=5$ and iv) blue curve for $\alpha=10$.
}
\label{fig:1}
\end{figure}

\begin{figure}[ht!]
\centering
\includegraphics[width=0.49\textwidth]{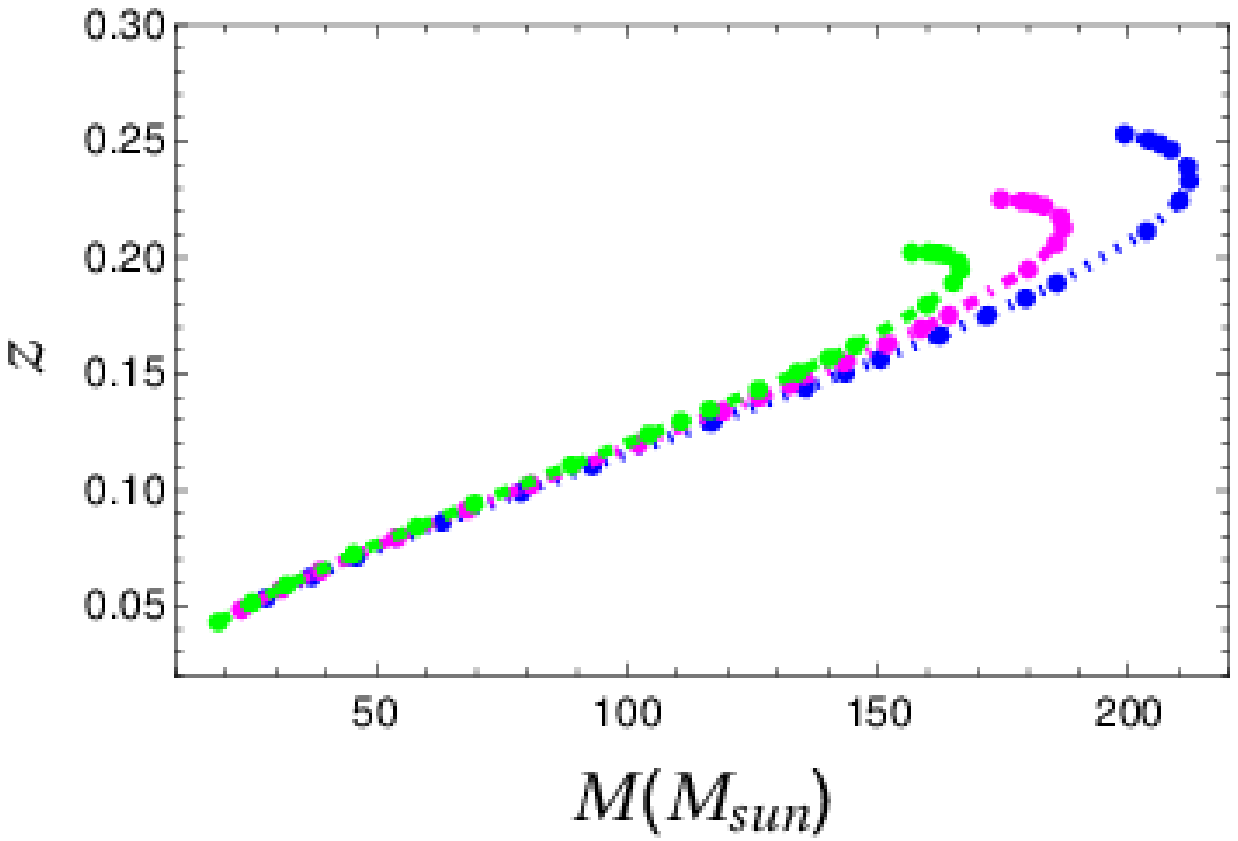} \ 
\includegraphics[width=0.49\textwidth]{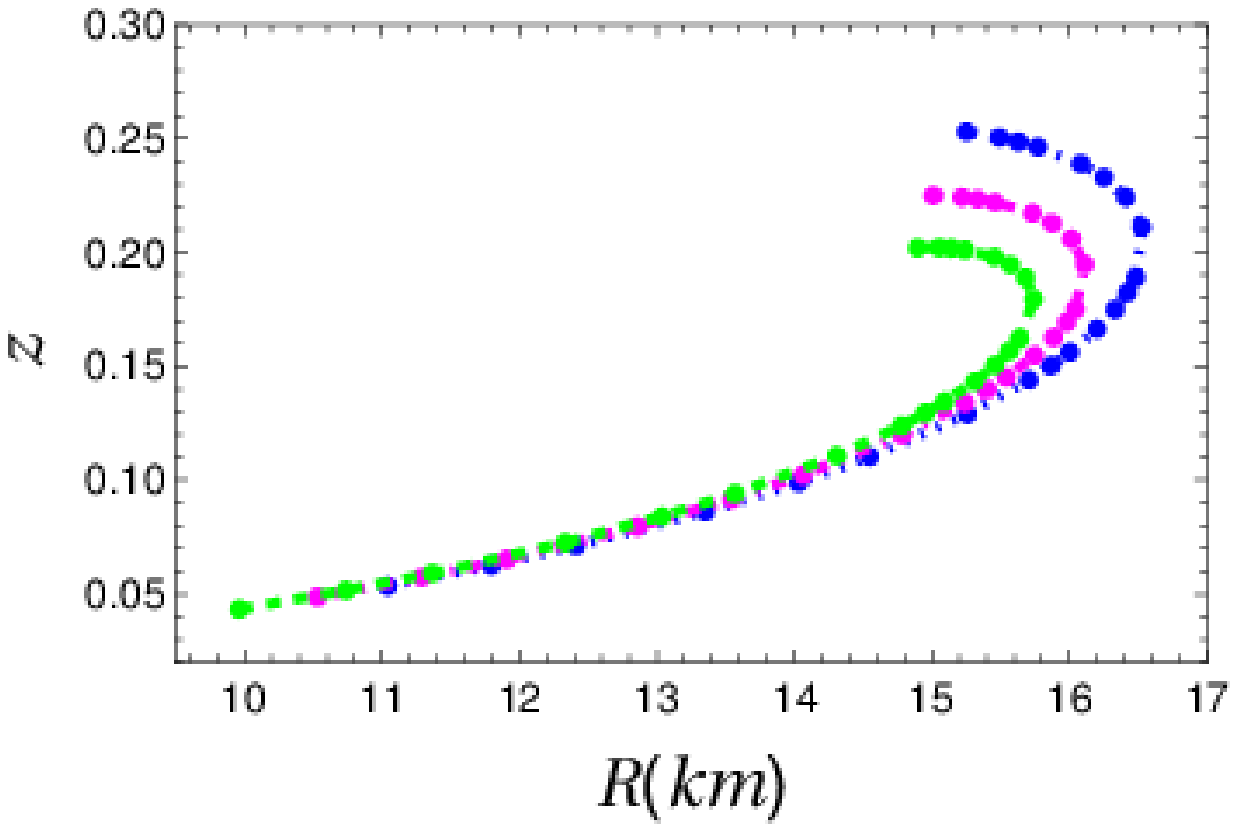}
\caption{{\bf Left panel:} 
Gravitational red-shift vs mass (in solar masses) of the star for three different values of the Gauss-Bonnet parameter. {\bf Right panel:} Gravitational red-shift vs radius (in km) of the star for three different values of the Guass-Bonnet parameter. The color code is as follows: i) green curve for $\alpha=1$, ii) magenta curve for $\alpha=5$ , iii) blue curve for $\alpha=10$.
}
\label{fig:2}
\end{figure}

\section{Numerical results}

\begin{table}
\centering
  \caption{Maximum values of star mass (in solar masses) and radius (in km)}
  \begin{tabular}{ccccc}
  \hline
                & $\alpha = 0$ & $\alpha = 1$ & $\alpha = 5$ & $\alpha = 10$  \\
\hline
$R_{\text{max}}$ &  15.5       &   15.7       &  16.1        &   16.5  \\
$M_{\text{max}}$ & 161.2       &  166.5       & 187.0        &  212.2  \\
\hline
\end{tabular}
\label{tab:ppnfr}
\end{table}

The model is characterized by 3 free parameters in total, namely $G_5, \alpha$ and $B$. The fundamental five-dimensional Planck mass, $8 \pi G_5 = M_*^{-3}$, is different than the usual four-dimensional one, $M_p = 2.4 \times 10^{18}$~GeV, and a priori it is unknown. The same holds for the other two free parameters $B, \alpha$.

In the discussion to follow, in order to obtain numerical solutions to the structure equations we will fix $B$ and $M_*$, and we shall investigate the impact of the Gauss-Bonnet coupling $\alpha$ on the the properties of strange quark stars in $1+4$ dimensions. The five-dimensional Planck mass in principle can take any value between the TeV scale and the usual four-dimensional Planck scale. A string scale/higher-dimensional Planck mass of the order of a few TeV is very attractive from the theoretical point of view for several reasons, as it addresses the hierarchy problem of Particle Physics, it provides an alternative to gauge coupling unification in D-brane constructions of the Standard Model \cite{kiritsis}, and it may explain anomalies related to cosmic ray observations \cite{tomaras}. Finally, the evaporation via the Hawking radiation of TeV mini-black holes can be seen at the colliders \cite{casanova}. Therefore, it is advantageous to consider a $M_*$ not too far away from, say, $10~TeV$, and in the rest of the article we shall assume that $M_* \sim 10^6$~GeV. What is more, we fix the bag constant to take the numerical value $B=1.6 \times 10^{-53}$. Then, the only free parameter left is the Gauss-Bonnet parameter (plus of course the central pressure, which is also unknown).

In the left panel of Fig.~\ref{fig:1} we show the mass-to-radius profiles for three different values of the Gauss-Bonnet parameter, $\alpha=1,5,10$, in units of the five-dimensional Planck mass. The case $\alpha=0$ corresponding to 5D GR is also shown for comparison reasons. We see that we obtain the usual profiles familiar from the standard four-dimensional GR, although the 5D objects are larger and much heavier than their 4D counterparts \cite{prototype,greg1}. First, the radius acquires a maximum value, $R_{\text{max}}$, while the star mass still increases. Then the mass, too, acquires a maximum value, $M_{\text{max}}$. Both $R_{\text{max}}$ and $M_{\text{max}}$ for all cases are shown in Table I. 

In the right panel of Fig.~\ref{fig:1} we show the compactness of the star as a function of the star mass. Since in the $\alpha \rightarrow 0$ limit we recover the Schwarzschild solution of five-dimensional GR
\begin{equation}
f(r)^{\alpha=0} = 1 - \bigg(\frac{1}{6 \pi^2 } \bigg)\frac{2M}{r^2}
\end{equation}
we define the compactness of the five-dimensional star to be 
\begin{equation}
\beta_5 = \bigg(\frac{1}{6 \pi^2}\bigg)  \frac{M}{R^2}
\end{equation}
in close analogy to the usual four-dimensional case, $\beta_4=M/R$. With that definition the compactness of a black hole is still 1/2, while any compact object has a compactness lower than that. 

Our results show that as we increase the Gauss-Bonnet parameter the stars become even more compact. Furthermore, for light stars the Gauss-Bonnet parameter has no observable effect. Only for sufficiently heavy stars, $M \geq 125~M_{\odot}$, the Gauss-Bonnet parameter starts to have a considerable impact on the profiles. Similar results were obtained in the four-dimensional Starobinsky model \cite{staro1,staro2}.

\begin{figure}[ht!]
\centering
\includegraphics[width=0.75\textwidth]{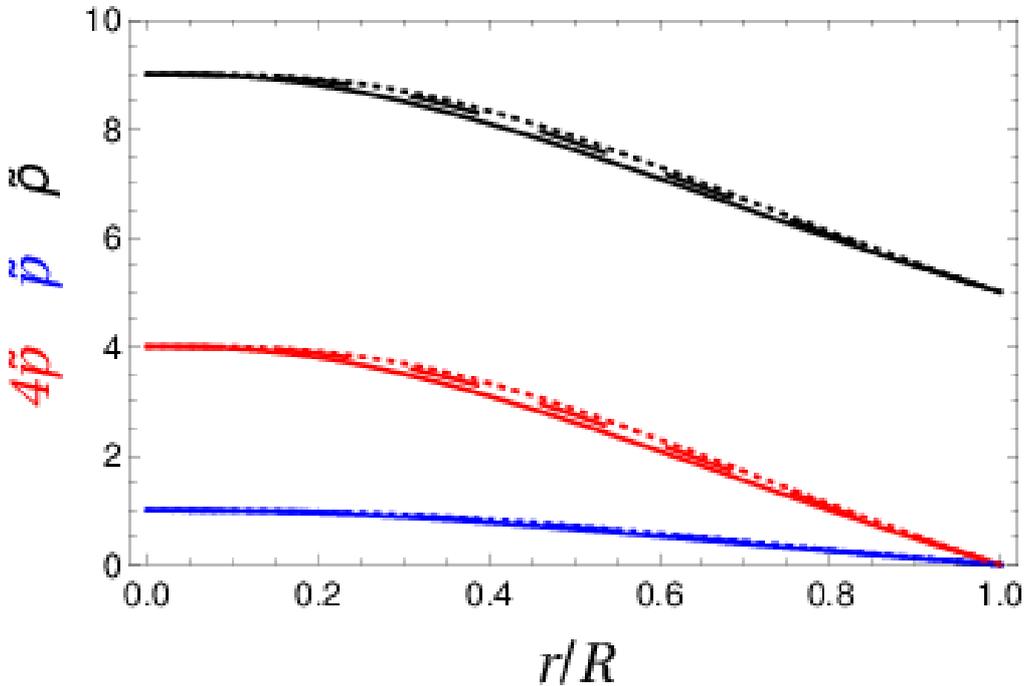}
\caption{
Checking the energy conditions for isotropic stars for $p_c(0)=B$ and $\alpha=1, 5, 10$. Shown are: Normalized energy density (black curves), normalized pressure (blue curves), and four times the normalized pressure (red curves). We have considered three different cases: i) $\alpha =1$ (solid line), ii) $\alpha = 5$ (long dashed line) and iii) $\alpha = 10$ (short dashed line).
}
\label{fig:3}
\end{figure}

The obtained solution should be able to describe realistic astrophysical configurations. Therefore, in this subsection as a final check we investigate if the energy conditions are fulfilled or not. We require that \cite{Ref_Extra_1,Ref_Extra_2,Ref_Extra_3,ultimo}
\begin{equation}
\rho \geq 0
\end{equation}
\begin{equation}
\rho - p  \geq  0
\end{equation}
\begin{equation}
\rho - 4 p \geq 0
\end{equation}
which are the five-dimensional version of the energy conditions.
Fig.~\ref{fig:3} shows the normalized pressure, $\tilde p \equiv p/B$, $4 \tilde p$, and normalized energy density, $\tilde \rho \equiv \rho/B$ (from bottom to top) versus the normalized radial coordinate $r/R$ for $\alpha=1, 5, 10$ and $p_c(0)=B$. We thus conclude that the solutions obtained in the present work are realistic solutions, which are able to describe realistic astrophysical configurations.

Finally, if the wavelength of a photon emitted at the surface of the star is $\lambda_e$, and the wavelength of the same photon observed at infinity is $\lambda_o$, the gravitational red-shift $z$ is defined to be the fractional variation of the wavelength
\begin{equation}
z = \frac{\lambda_o - \lambda_e}{\lambda_e} = -1 + \frac{\lambda_o}{\lambda_e} = -1 + \frac{\omega_e}{\omega_o}
\end{equation}
where now $\omega_e$ and $\omega_o$ are the emitted frequency and the observed frequency of the photon, respectively. The ratio of the two is given by \cite{prototype}
\begin{equation}
\frac{\omega_e}{\omega_o} = \left( \frac{f(\infty)}{f(R)} \right)^{1/2}
\end{equation}
where the metric function is evaluated at the surface of the star and at infinity. For asymptotically flat spacetimes $f(\infty) = 1$, and therefore the gravitational red-shift is given by the final expression \cite{angel}
\begin{equation}
z = -1 + \frac{1}{\sqrt{f(R)}}
\end{equation}
In Fig.~\ref{fig:2} we show $z$ vs the star mass (left panel) as well as vs the radius of the star  (right panel) for three different values of the Gauss-Bonnet parameter, $\alpha=1,5,10$ in units of the five-dimensional Planck mass.

Clearly, the present work could be extended in various directions. 
It would be interesting to study, for instance, rotating stationary axisymmetric solutions, strange quark stars using a more sophisticated EoS, neutron stars using an accurate, high-quality EoS, and oscillations of compact objects, to mention just a few. In addition, it would be worth-pursuing investigating the impact of
the extra dimension and the Gauss-Bonnet term on properties of compact
objects in the brane-world scenario 
\cite{Randall:1999ee,Randall:1999vf,Dvali:2000hr}  
using the effective field
equations on the brane. In that framework, in order to avoid
conflicts will well-tested laws, such as Coulomb's and Newton's laws, ordinary matter is not allowed to propagate into the bulk, and any astrophysical object is therefore required to be confined on the brane.
What is more, it would be interesting to investigate how the properties of the solution obtained here are modified in the framework of the so-called scale-dependent scenario, where the coupling constants acquire a dependence on the scale, i.e.  $\{G_0,\Lambda_0\} \rightarrow \{G_k, \Lambda_k\}$), and which has received considerable attention lately
\cite{Koch:2016uso,Rincon:2017ypd,Rincon:2017goj,Rincon:2017ayr,
Contreras:2017eza,Hernandez-Arboleda:2018qdo,Contreras:2018dhs,Rincon:2018lyd,Rincon:2018sgd,Contreras:2018swc,Contreras:2018gpl,Contreras:2018gct,Canales:2018tbn,Rincon:2019cix,Rincon:2019zxk,Contreras:2019fwu,Fathi:2019jid}. In this case, the TOV equations should be modified to account for the running of the gravitational coupling and the Gauss-Bonnet parameter $\alpha$. We hope to be able to address some of these issues in future works.

\section{Anisotropic objects}

In studies of compact relativistic astrophysical objects the authors usually focus on stars made of isotropic fluids, where the radial pressure $p_r$ equals the tangential pressure $p_T$. Celestial bodies, however, under certain conditions may become anisotropic. Such a possibility was mentioned for the first time in~\cite{aniso1}, where the author made the observation that relativistic particle interactions in a very dense nuclear matter medium could lead to the formation of anisotropies. Indeed, anisotropies arise in many scenarios of a dense matter medium, like phase transitions \cite{aniso3}, pion condensation \cite{aniso4}, or in presence of type $3A$ super-fluid \cite{aniso5}. 

In the case of an anisotropic star we define the anisotropy $\Delta \equiv p_T - p_r$, and the continuity equation becomes 
\begin{equation}
p_r'(r) =  - (p_r(r) + \rho(r)) \: \frac{\frac{p_r(r) r^3}{3} + r (1-e^{-2 \lambda})}{(r^2 + 4\alpha) e^{-2 \lambda} - 3 \alpha e^{-4 \lambda}} + \frac{3 \Delta}{r}
\end{equation}
Clearly, for isotropic objects $\Delta=0$ and one recovers the usual continuity equation of the previous section.
 
Since now there is one more unknown quantity, to obtain a tractable solution an approach often used in the literature is to assume a concrete and reasonable mass function, see e.g. \cite{anisotropy,ultimo}. Then the rest of the quantities may de computed one by one using the EoS as well as the structure equations. Following~\cite{ultimo} we assume that
\begin{equation}
2 \lambda(r) = A r^2 = a (r/r_0)^2
\end{equation}
with $r_0$ being a length scale, and $a$ being a dimensionless number. The mass function $m(r)$ is computed to be
\begin{equation}
m(r) = 3 \pi^2 e^{-a(r/r_0)^2} (1-e^{-a(r/r_0)^2}) (r^2+2 \alpha-2 \alpha e^{-a(r/r_0)^2})
\end{equation}
The energy density is computed using the first structure equation
\begin{equation}
\rho(r) = \frac{m'(r)}{2 \pi^2 r^3}
\end{equation}
while $p_r$ is computed using the EoS
\begin{equation}
p_r(r) = \frac{1}{4} \: (\rho(r)-5B)
\end{equation}
Finally, the anisotropy is computed using the above continuity equation. 

\begin{figure}[ht!]
\centering
\includegraphics[width=0.32\textwidth]{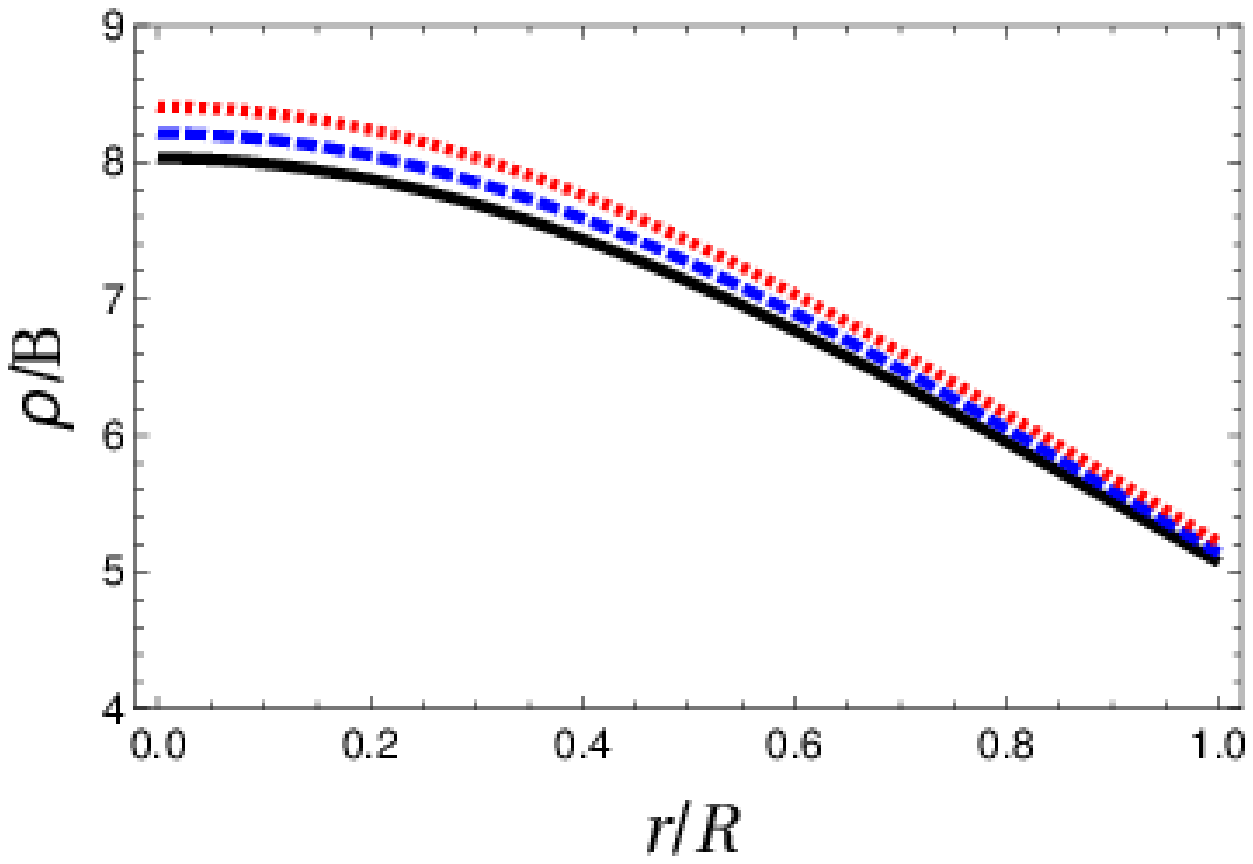} \ 
\includegraphics[width=0.32\textwidth]{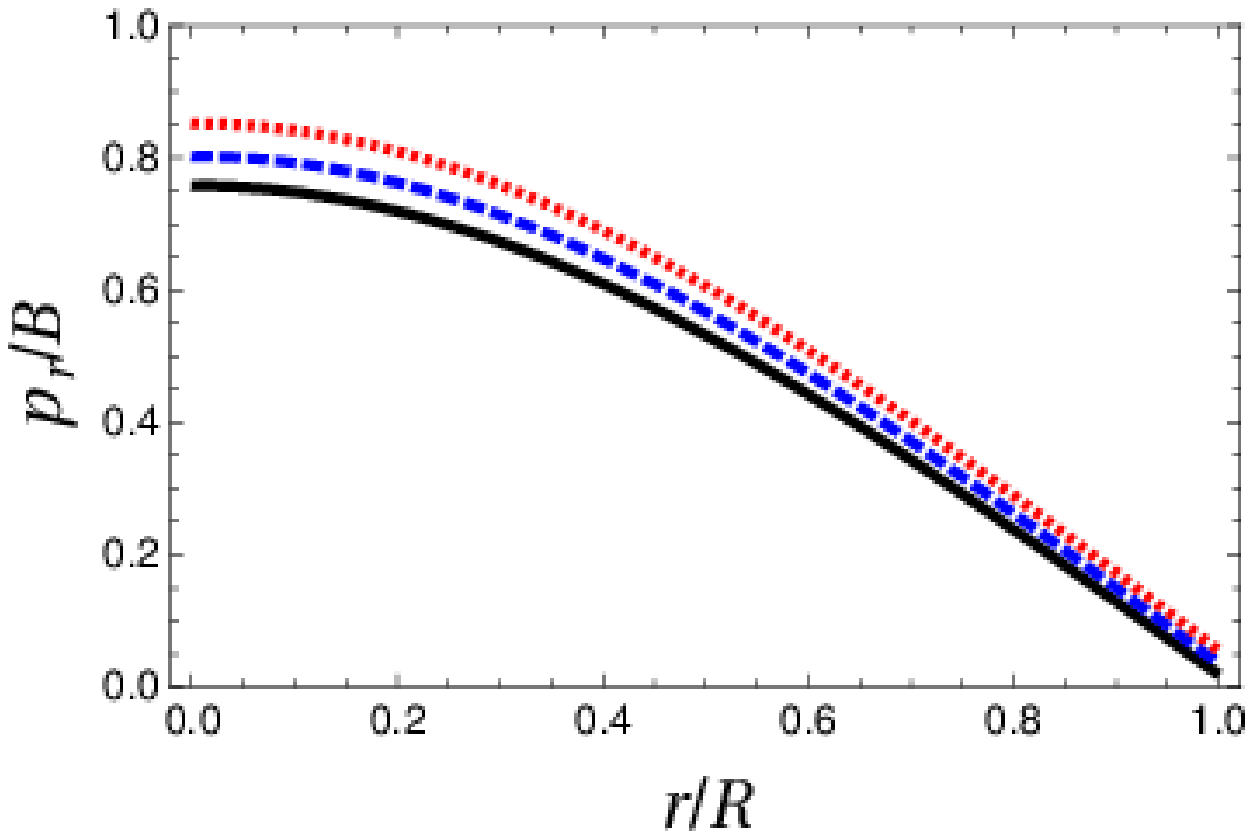} \
\includegraphics[width=0.32\textwidth]{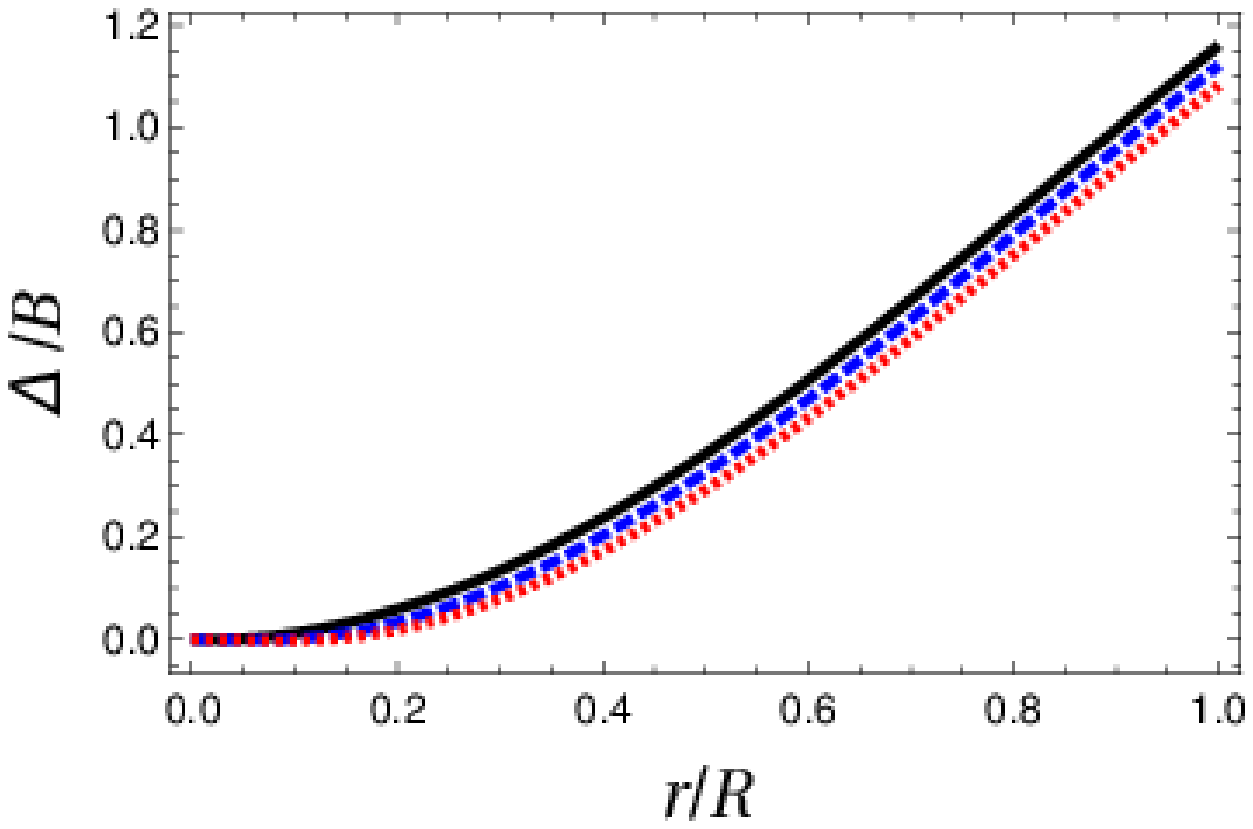}
\caption{
Interior solutions for anisotropic stars for i) $\alpha=1$ (solid black line), ii) $\alpha=10$ (dashed blue line), iii) $\alpha= 20$ (dotted red line) for $a=2.5$ and $r_0=45~km$.
{\bf Left panel:} Energy density versus normalized radial coordinate. 
{\bf Middle panel:} Radial pressure versus normalized radial coordinate.
{\bf Right panel:} Anisotropy versus normalized radial coordinate.
}
\label{fig:4}
\end{figure}

In Fig.~\ref{fig:4} we show the (normalized quantities) energy density, the radial pressure and the anisotropy for $\alpha=1, 10, 20$, $a=2.5$ and $r_0=45~km$.

We see that in all three cases the anisotropy vanishes at the centre of the star, and then it increases towards the surface. Although the radius of the strange quark star is computed demanding that $p_r=0$ at the surface of the object, the anisotropy is not required to vanish at the surface of the star. In addition, the anisotropy decreases as the Gauss-Bonnet parameter increases. The same impact was observed previously in \cite{ultimo}. The overall behaviour of the internal solution for the anisotropic strange quark stars is qualitatively similar to the one found in \cite{ultimo}.

\section{Conclusions}

To summarize, in the present work we have studied relativistic non-rotating stars in the framework of five-dimensional Lovelock gravity with the Gauss-Bonnet term. Matter inside the star is modelled as a relativistic gas of de-confined quarks, and we have assumed the simplest version of the MIT bag model "radiation plus constant". The impact of the Gauss-Bonnet parameter on properties of the stars has been investigated. We have integrated the modified Tolman-Oppenheimer-Volkoff equations numerically, and we have obtained the mass-to-radius profiles, the compactness of the stars as well as the gravitational red-shift for three different values of the Gauss-Bonnet parameter. Our results show that i) the stars become more and more compact as we increase the Gauss-Bonnet parameter, and ii) the latter has an observable effect only for sufficiently heavy stars, while for light stars there is no considerable impact on the mass-to-radius profiles. The maximum star mass and radius are also reported. The impact of the Gauss-Bonnet coupling on the interior solution of anisotropic quark stars has been studied as well.

\section{Acknowledgments}

We wish to thank the anonymous reviewer for useful suggestions. The author G.~P. thanks the Funda\c c\~ao para a Ci\^encia e Tecnologia (FCT), Portugal, for the financial support to the Center for Astrophysics and Gravitation-CENTRA, Instituto Superior T\'ecnico, Universidade de Lisboa, through the Grant No. UID/FIS/00099/2013.
The author A.~R. acknowledges DI-VRIEA for financial support through Proyecto Postdoctorado 2019 VRIEA-PUCV.


\end{document}